\documentclass{PoS}

\usepackage{bbm}

\newcommand{\cO}{\cal{O}}
\newcommand{\vs}{\vspace}

\newcommand{\bdm}{\begin{displaymath}}
\newcommand{\edm}{\end{displaymath}}
\newcommand{\beq}{\begin{equation}}
\newcommand{\eeq}{\end{equation}}
\newcommand{\bea}{\begin{eqnarray}}
\newcommand{\eea}{\end{eqnarray}}
\newcommand{\bit}{\begin{itemize}}
\newcommand{\eit}{\end{itemize}}
\newcommand{\bc}{\begin{center}}
\newcommand{\ec}{\end{center}}
\newcommand{\re}{\relax{\rm I\kern-.18em R}}
\newcommand{\D}{{\cal D}^{(ov)}}
\newcommand{\SD}{\hat{\cal D}^{(ov)}}
\newcommand{\Dinv}{\left[{\cal D}^{(ov)}\right]^{-1}}

\newcommand{\SDw}{\hat{\cal D}^{(W)}}
\newcommand{\ImpSpace}{{\cal P}}

\newcommand{\sumFL}{\sum\limits_{i=1}^{N_f}}

\newcommand{\ID}{\mathbbm{1}}

\newcommand{\ie}{{\it i.e. }}

\title{On the phase structure of a chiral invariant Higgs-Yukawa model}

\ShortTitle{On the phase structure of a chiral invariant Higgs-Yukawa model}

\author{\speaker{Philipp Gerhold}\\
        Institut f\"ur Physik, Humboldt Universit\"at zu Berlin, D-12489 Berlin, Germany\\
        E-mail: \email{gerhold@physik.hu-berlin.de} }

\author{Karl Jansen\\
        John von Neumann Institute for computing, NIC, \\
        Platanenallee 6, D-15738 Zeuthen, Germany\\
        E-mail: \email{Karl.Jansen@desy.de}}

\abstract{
In the past the construction of Higgs-Yukawa models on the lattice was blocked by the
lack of a consistent definition of a chiral invariant Yukawa coupling term.
Here, we consider a chiral invariant Higgs-Yukawa model based on the overlap operator $\D$
realized by the Neuberger-Dirac operator. As a first step towards a numerical
examination of this model we study its phase diagram by means of an analytic $1/N_f$-expansion,
which is possible for small and for large values of the Yukawa coupling constant. In the case 
of strong Yukawa couplings the model effectively becomes an $O(4)$-symmetric non-linear 
$\sigma$-model.}

\FullConference{XXIVth International Symposium on Lattice Field Theory\\
                July 23-28, 2006\\
                Tucson, Arizona, USA}

\begin{document}

\section{Introduction}
Non-perturbative investigations of Higgs-Yukawa models using the lattice regularization became subject 
of many investigations in the early 1990's~\cite{Smit:1989tz,Shigemitsu:1991tc,Golterman:1990nx,book:Montvay,book:Jersak,Golterman:1992ye}.
These lattice studies were initially motivated by the interest in a better understanding of the fermion mass 
generation via the Higgs mechanism on a non-perturbative level and, in particular, in the determination of bounds
on the Yukawa couplings translating into bounds on the Higgs boson mass and the - at that time not yet
discovered - top quark mass. Furthermore, the fixed point structure of the theory received special attention
due to the question whether besides the Gaussian also a non-trivial fixed point might exist. However, these 
investigations were blocked, since the influence of unwanted fermion doublers could not successfully be 
suppressed. Moreover, the models of these studies suffered the lack of chiral symmetry. The latter, 
however, would be indispensable for a consistent lattice regularization of chiral gauge theories like, for 
example, the standard model of electroweak interactions. 

Here, we follow the proposition of L\"uscher~\cite{Luscher:1998pq} for a chiral invariant lattice Higgs-Yukawa 
model based on the Neuberger overlap operator~\cite{Neuberger:1998wv}. As a first step we begin with an 
analytical investigation of its phase structure by means of $1/N_f$-expansions following \cite{Hasenfratz:1991it,Hasenfratz:1992xs}
and the references therein. We derive an expression for the effective potential at tree-level and present our preliminary 
results for the corresponding phase diagram.

In chapter~\ref{chap:model} we briefly describe the considered model, before we present the large $N_f$ results at
tree-level for small Yukawa couplings $y_N\propto1/\sqrt{N_f}$ in chapter~\ref{chap:SmallYukawa}. In the following 
chapter~\ref{chap:FiniteYukawa} we discuss a different large $N_f$-limit which becomes valid for non-vanishing Yukawa 
couplings. We then end with a short outlook.

\section{The model}
\label{chap:model}
The chiral invariant Higgs-Yukawa model, which we consider here, contains one four-component,
real Higgs field $\Phi$ and $N_f$ fermion doublets represented by eight-component spinors $\psi^{(i)}$,
$\bar\psi^{(i)}$ with $i=1,...,N_f$. Furthermore, there are also $N_f$ {\it auxiliary} fermionic doublets 
$\chi^{(i)}$, $\bar\chi^{(i)}$ only introduced to construct a chiral invariant Yukawa interaction term. 
The partition function can then be written as 
\beq
Z = \int D\Phi\,\prod\limits_{i=1}^{N_f} \left[D\psi^{(i)}\, D\bar\psi^{(i)}\, D\chi^{(i)}\,
D\bar\chi^{(i)} \right]\,  \exp\left( -S_\Phi -S_F^{kin} - S_Y  \right)
\eeq
where the total action is decomposed into the Higgs action $S_\Phi$, the kinetic fermion action $S_F^{kin}$, 
and the Yukawa coupling term $S_Y$. It should be stressed that no gauge fields are considered here. 

The kinetic fermion action describes the propagation of the physical fermion fields $\psi^{(i)}$,$\bar\psi^{(i)}$ 
in the usual way according to 
\bdm
S_F^{kin} = \sumFL \sum\limits_{n,m} \bar\psi^{(i)}_n \D_{n,m} \psi^{(i)}_{m} -
2\rho\bar\chi^{(i)}_n\ID_{n,m} \chi^{(i)}_m
\edm
where the coordinates $n,m$ as well as all field variables and coupling constants are given in lattice units
throughout this paper. The (doublet) Dirac operator $\D= \SD\otimes\SD$ is given by the Neuberger overlap operator 
$\SD$, which is related to the Wilson operator $\SDw=\gamma^E_\mu \nabla_\mu - r \nabla_\mu\nabla_\mu$ by 
\bea
\SD &=& \rho\left\{1+\frac{\hat A}{\sqrt{\hat A^\dagger \hat A}}   \right\},\quad \hat A = \SDw - \rho, \quad \rho \ge 1
\eea
where $\nabla_\mu$ denotes the symmetric difference quotient. It is well known that the eigenvalues $\nu^\pm(p)$ of $\SD$
with $\mbox{Im}[\nu^\pm(p)] \gtrless0$ form a circle in the complex plane, the radius of which is given by the 
parameter $\rho$. In momentum space with $p \in \ImpSpace=[-\pi,\pi]^{\otimes 4}$ these eigenvalues are explicitly
given by
\bea
\nu^\pm(p)&=& \rho + \rho\cdot\frac{\pm i\sqrt{\tilde p^2} + r\hat p^2 - \rho}{\sqrt{\tilde p^2 + (r\hat p^2 -
\rho)^2}},\quad \tilde p_\mu = \sin(p_\mu),\quad \hat p_\mu = 2 \sin\left(\frac{p_\mu}{2}\right).
\eea
The auxiliary fields $\chi^{(i)}$ on the other hand do not propagate. They do not have a direct physical 
interpretation at all and their contribution to the action is only introduced to establish chiral symmetry.

The Higgs field couples to the fermions according to the Yukawa coupling term
\beq
S_Y = y_N \sum\limits_{n,m}\sumFL(\bar\psi^{(i)}_n+\bar\chi^{(i)}_n) \underbrace{\left[
\ID_{n,m}\frac{(1-\gamma_5)}{2}\phi_n 
+ \ID_{n,m}\frac{(1+\gamma_5)}{2}\phi^{\dagger}_n  \right]}_{B_{n,m}} (\psi^{(i)}_m+\chi^{(i)}_m)
\label{eq:DefYukawaCouplingTerm}
\eeq
where $y_N$ denotes the Yukawa coupling constant and $B_{n,m}$ will be referred to as Yukawa coupling matrix. 
Here the Higgs field $\Phi_n$ is rewritten as a quaternionic, $2 \times 2$ matrix 
$\phi_n = \Phi_n^0\ID -i\Phi_n^j\tau_j$ ($\tau_j$: Pauli matrices) acting on the flavor index. 
Due to the chiral character of this model, left- and right-handed fermions couple differently to the 
Higgs field, as can be seen from the appearance of the projectors $(1\pm \gamma_5)/2$ in the Yukawa term.

Finally, we use a slightly unusual notation for the Higgs action $S_\Phi$ given by
\beq
\label{eq:ModPhiAction}
S_\Phi = -\kappa_N\sum_{n,\mu} \Phi_n^{\dagger} \left[\Phi_{n+\hat\mu} + \Phi_{n-\hat\mu}\right]
+ \sum_{n} \Phi^{\dagger}_n\Phi_n + \lambda_N \sum_{n} \left(\Phi^{\dagger}_n\Phi_n - N_f \right)^2
\eeq
where $\kappa_N$ denotes the hopping parameter and $\lambda_N$ is the quartic coupling. The usual notation
is reobtained by a trivial rescaling of the coupling constants and the Higgs field. The model then obeys an exact, but 
{\it lattice modified} chiral symmetry recovering the actual chiral symmetry in the continuum limit~\cite{Luscher:1998pq}.

For the further analytical treatment of this model the fermionic degrees of freedom are
integrated out leading to 
the effective action $S_{eff}[\Phi]$, which can be written in terms of fermionic determinants
yielding
\beq
\label{eq:effectiveHiggsAction1}
S_{eff}[\Phi]= S_\Phi[\Phi] - N_f\cdot \log\left[\det\left( y_NB\D -2\rho\D -2\rho y_N B \right) \right].
\eeq

\section{Large $N_f$-limit for small $y_N$}
\label{chap:SmallYukawa}
We now consider the limit of infinite fermion number $N_f \rightarrow \infty$ with the coupling constants scaling 
according to
\bea
\label{eq:LargeNBehaviourOfCouplings1}
y_N = \frac{\tilde y_N}{\sqrt{N_f}},\; \tilde y_N=\mbox{const}\quad & \lambda_N = \frac{\tilde \lambda_N}{N_f},\; \tilde \lambda_N=\mbox{const}&
\quad\kappa_N = \tilde \kappa_N,\; \tilde\kappa_N = \mbox{const} 
\eea
which justifies a semi-classical approach to the investigated model. We will therefore directly evaluate the
effective action (\ref{eq:effectiveHiggsAction1}), which is at least possible for the constant and the staggered mode
of the Higgs field. Since the Higgs field scales proportional to $\sqrt{N_f}$ in the large $N_f$-limit, we apply
the ansatz
\beq
\label{eq:staggeredAnsatz}
\Phi_n = \hat\Phi \cdot \sqrt{N_f} \cdot \left(m + s\cdot (-1)^{\sum\limits_{\mu}n_\mu}    \right) 
\eeq
where $\hat\Phi\in \re^4$ with $|\hat\Phi|=1$ denotes a constant 4-dimensional unit vector. We will refer to $m$, 
$s$ as magnetization and staggered magnetization, respectively. For the actual evaluation of the effective action 
we rewrite (\ref{eq:effectiveHiggsAction1}), neglecting all constant terms independent of $\Phi$ in the following, as
\bea
S_{eff}[\Phi]&=& S_\Phi[\Phi] - N_f\cdot \log\left[\det\left(\ID - \frac{y_N}{2\rho}\cdot\left(\D-2\rho\right)\cdot\Dinv\cdot B \right) \right].
\label{eq:EffActionRewrittenForSmallY}
\eea
For the effective potential $V_{eff}(m,s)$ at tree-level one then finally finds
\bea
\frac{1}{N_f}V_{eff}\left(m,s\right)   
&=& -\tilde\kappa_N \Big(m^2-s^2\Big) +  m^2+s^2 
+\tilde\lambda_N \Big( m^4 +s^4 + 6m^2s^2 -2\left(m^2+s^2 \right) \Big)   \nonumber\\   
&-& \int\limits_{p\in\ImpSpace} \frac{d^4p}{(2\pi)^4}\,\log\Bigg[ \left(1+\left(\frac{\tilde y_N}{2\rho} \right)^2 \left(m^2-s^2\right)
\frac{|\nu(p)-2\rho|}{|\nu(p)|}  \cdot  \frac{|\nu(\wp)-2\rho|}{|\nu(\wp)|}  \right)^2 \nonumber\\
&+& m^2 \left(\frac{\tilde y_N}{2\rho} \right)^2 \left(\frac{|\nu(p)-2\rho|}{|\nu(p)|} - \frac{|\nu(\wp)-2\rho|}{|\nu(\wp)|}   \right)^2 \Bigg]
\eea
with the abbreviations $\wp_\mu = p_\mu + \pi$.

The phase diagram can then be explored numerically by searching for the absolute minima
of the effective potential $V_{eff}(m,s)$ with respect to $m$ and $s$. In general, four types of 
solutions can be obtained. These are a symmetric (SYM: $m=0$, $s=0$), a ferromagnetic 
(FM: $m\neq 0$, $s=0$), an antiferromagnetic (AFM: $m=0$, $s\neq 0$), and a ferrimagnetic phase (FI: $m\neq 0$, $s\neq 0$).
The corresponding phase diagram for the choice $\tilde\lambda_N=0.1$ is shown in Figure~\ref{fig:PhaseDiagram1}.
Here, we only present a qualitative plot of the phase structure, since the obtained results are preliminary.
However, the diagram reveals a rich phase structure. Especially, there is a ferrimagnetic phase, which was also 
observed in Monte Carlo studies of earlier, not chiral invariant Higgs-Yukawa models, for example in~\cite{Bock:1990tv}.
\bc
\setlength{\unitlength}{0.01mm}
\begin{figure}[htb]
\centering
\begin{picture}(5700,5100)
\put(300,100){\includegraphics[width=5cm]{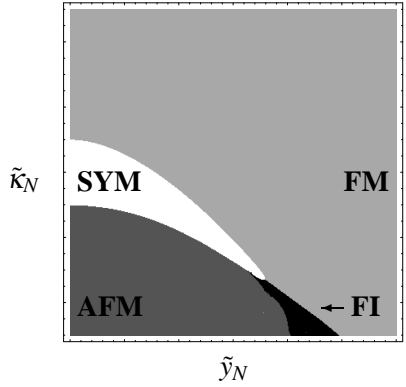}}
\put(25,2500){$\tilde \kappa_N$}
\put(2850,0){$\tilde y_N$}
\put(950,2450){\textbf {SYM}}
\put(4500,2450){\textbf {FM}}
\put(950,800){\textbf {AFM}}
\put(4600,800){\textbf {FI}}
\put(4500,900){\vector(-1,0){300}}
\end{picture}
\caption{Qualitative phase diagram with respect to the Yukawa coupling $\tilde y_N$ and the hopping parameter $\tilde \kappa_N$
for a constant quartic coupling $\tilde \lambda_N=0.1$.}
\label{fig:PhaseDiagram1}
\end{figure}
\ec
\vs{-5mm}

\section{Large $N_f$-limit for finite $y_N$}
\label{chap:FiniteYukawa}
\begin{figure}[htb]
\setlength{\unitlength}{0.01mm}
\mbox{
\begin{minipage}{0.5\linewidth}
\begin{picture}(6000,5400)
\put(1400,400){\includegraphics[height=5cm]{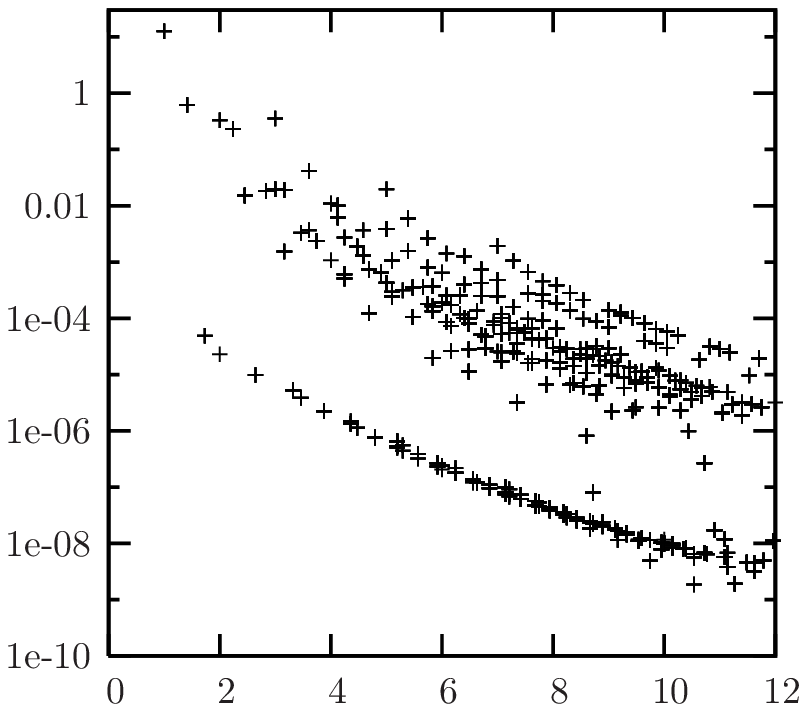}}
\put(0,3100){$|K(\Delta x)|$}
\put(4150,70){$|\Delta x|$}
\end{picture}
\end{minipage}

\begin{minipage}{0.5\linewidth}
\begin{picture}(5800,5400)
\put(800,200){\includegraphics[height=5.4cm]{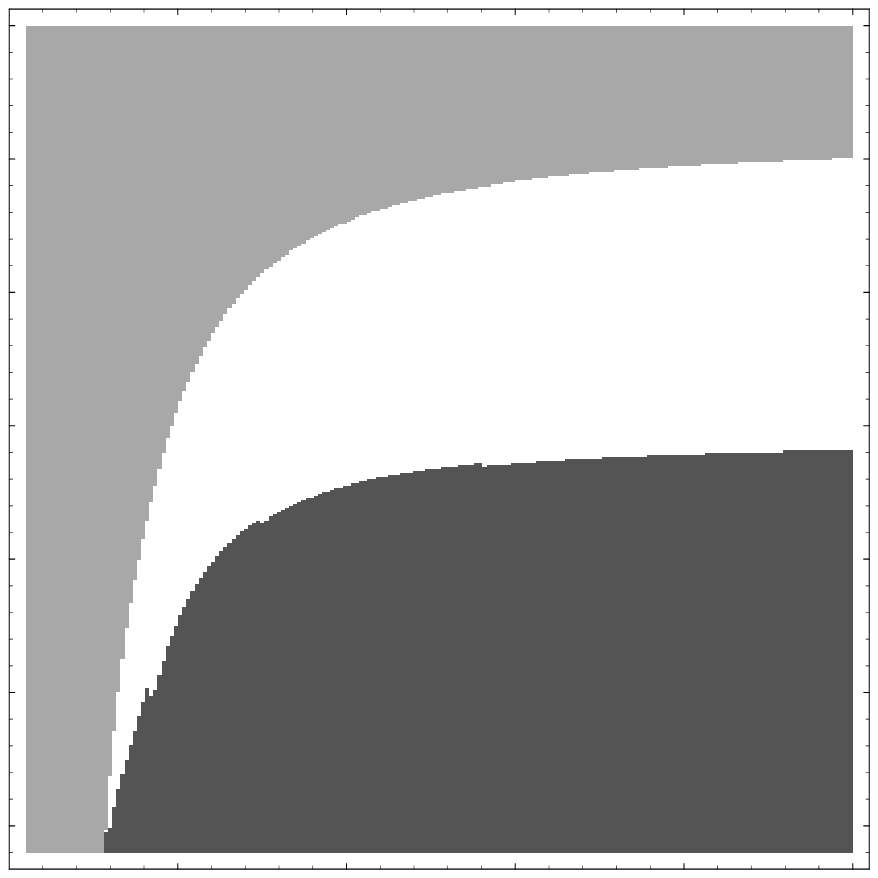}}
\put(700,3100){$\tilde \kappa_N$}
\put(3400,70){$\tilde y_N\gg 1$}
\put(4000,1600){\textbf {AFM}}
\put(1800,3500){\textbf {FM}}
\put(4000,3500){\textbf {SYM}}
\put(1680,600){\textbf {//}}
\end{picture}
\end{minipage}
}
\caption{Left: Square-norm of the coupling matrix $K(\Delta x)$ versus the square-norm of the distance $\Delta x$ in lattice units.
Right: Qualitative phase diagram with respect to the Yukawa coupling $\tilde y_N\gg 1$ and the hopping parameter $\tilde \kappa_N$
for a constant quartic coupling $\tilde\lambda_N=0.1$.}
\label{fig:PhaseDiagram2}
\end{figure}

We now consider a different limit of large fermion numbers where the coupling constants scale according to
\bea
\label{eq:LargeNBehaviourOfCouplings2}
y_N = \tilde y_N,\; \tilde y_N=\mbox{const}\quad & \lambda_N = \frac{\tilde \lambda_N}{N_f},\; \tilde \lambda_N=\mbox{const}&
\quad\kappa_N = \frac{\tilde \kappa_N}{N_f},\; \tilde\kappa_N = \mbox{const}.
\eea
Again, the Higgs field scales proportional to $\sqrt{N_f}$ allowing to expand the effective action in powers of 
$1/(y_N|\Phi|)$. In this large $N_f$-limit the power series expansion 
\bea
S_{eff}[\Phi] &=& - N_f\cdot \left( Tr\,\log\left[B  \right] 
- \sum\limits_{k=1}^{\infty} \frac{2^k}{k} \left(\frac{\rho}{y_N}\right)^k Tr\, \left[\D \left(\D-2\rho \right)^{-1} B^{-1}\right]^k \right)  \nonumber\\
&+& S_\Phi[\Phi] 
\eea
of the logarithm in equation~(\ref{eq:EffActionRewrittenForSmallY}) can therefore be cut off after the first 
non-vanishing term. The first summand $(k=1)$ is identical to zero and the second term $(k=2)$ 
is the first non-vanishing contribution. Cutting off the power series at $k=2$, the model then becomes 
an effective, $O(4)$-symmetric spin model
\bea
S_{eff}[\Phi] &=& S_\Phi[\Phi] 
- 4 N_f\cdot \left( 
\sum\limits_{n} \log\left(|\Phi_n|^2\right) +  
\frac{(2\rho)^2}{y_N^2}  \sum\limits_{n,m} \frac{\Phi_n^\mu}{|\Phi_n|^2} \cdot K_{\mu\nu}(\Delta x)\cdot
\frac{\Phi_m^\nu}{|\Phi_m|^2} \right) 
\eea
where $\Delta x = n-m$ and the non-local coupling matrix 
$K_{\mu\nu}(\Delta x)=\delta_{\mu,\nu} \cdot \left|\Gamma(\Delta x)  \right|^2$ 
with $|.|$ denoting the 4-vector norm
is explicitly given by the momentum integral
\bea
\re\ni \Gamma_\mu(\Delta x) &=& \int\limits_{p\in\ImpSpace} \frac{d^4p}{(2\pi)^4}e^{ip\Delta x} \cdot
\frac{\nu^+(p)}{\nu^+(p) - 2\rho} \cdot \frac{ \tilde p_\mu }{\sqrt{\tilde p^2}},
\eea
which can be computed numerically. One then finds that the square-norm of the coupling matrix, 
$|K(\Delta x)|= \left|\Gamma(\Delta x)  \right|^2$, decays exponentially with 
increasing distance $|\Delta x|$ as shown in Figure \ref{fig:PhaseDiagram2}.
In a field-theoretical sense the effective spin model therefore remains a locally interacting
model.

For the evaluation of the corresponding phase diagram of this effective spin model we make the ansatz
\beq
\Phi_n = \sqrt{N_f} \cdot \varphi_0 \cdot \sigma_n, \quad \sigma_n\in\re^4,\;|\sigma_n| = 1,\quad \re\ni\varphi_0 = \mbox{const}
\eeq
for the Higgs field. Considering only the leading power in $1/N_f$ of the tree-level effective action $S_{eff}[\Phi]$, 
the amplitude $\varphi_0$ can be fixed according to
\beq
0 = -4 \cdot \frac{1}{\varphi_0^2} + 1 + 2\tilde\lambda_N\cdot \left(\varphi_0^2-1 \right).
\eeq
Including the next to leading order terms in $1/N_f$, the model effectively becomes an $O(4)$-symmetric 
non-linear $\sigma$-model
\bea
S_{eff}[\Phi] &=&-\sum\limits_{n,m} \kappa^{eff}_{n,m} \cdot \sigma_n\cdot \sigma_m, \quad \mbox{with}  \\
\kappa^{eff}_{n,m} &=&  \frac{16\rho^2}{\tilde y_N^2 \varphi_0^2} \cdot \left|\Gamma(\Delta x)\right|^2 + 
\tilde\kappa_N\cdot \varphi_0^2\cdot\sum\limits_{\mu=\pm 1}^{\pm 4} \delta_{n, m+\hat\mu},
\label{eq:NonLinearSigmaModel}
\eea
the phase structure of which is again accessable to an $1/N$-expansion. Here, $N$ denotes the number of
components of the Higgs field $\Phi$, \ie eventually $N=4$. As usual, the constraint $|\sigma_n|=1$ is 
removed by the introduction of an auxiliary field $\lambda_n$ according to
\bea
Z &=& \int D\lambda\,\prod\limits_{i=1}^N\left[D\sigma^i\right]\, \exp\left[-S[\sigma,\lambda]  \right],\\
S[\sigma,\lambda] &=& \frac{1}{t_N}\cdot\left\{\sum\limits_{n,m}\sum\limits_{i=1}^N -\kappa^{eff}_{n,m} \cdot
\sigma^i_n\cdot \sigma^i_m + \sum\limits_n\lambda_n
\cdot\left[ \sum\limits_{i=1}^N\left(\sigma^i_n\right)^2 -1\right]\right\}.
\eea
Integrating out the $N-1$ components $\sigma^2$,...,$\sigma^N$ one obtains the reduced action 
$\hat S[\sigma^1,\lambda]$ depending only on the fields $\lambda_n$ and $\sigma^1_n$ given by
\bea
\hat S[\sigma^1,\lambda] &=& -\sum\limits_{n,m} \kappa^{eff}_{n,m}\cdot \frac{1}{t_N} \cdot \sigma^1_n\cdot \sigma^1_m + \sum\limits_{n}\lambda_n
\cdot \frac{1}{t_N}\cdot\left[ \left(\sigma^1_n\right)^2 -1\right]\nonumber\\
&+& \frac{1}{2}(N-1)\,\mbox{Tr}_{n,m}\,\log\left[-\kappa^{eff}_{n,m} + \lambda_n\delta_{n,m}  \right].
\eea
The newly introduced parameter $t_N$ is necessary to perform the large $N$-limit, which will be done
for $\tilde t_N \equiv t_N\cdot N = \mbox{const}$. Here, we are actually interested in the case $t_N=1$ and 
$N=4$. We therefore choose $\tilde t_N=4$. Again, we consider the magnetization and a possible staggered magnetization
for the $\sigma^1_n$ field and assume the auxiliary $\lambda_n$ field to be constant leading to the ansatz
\bea
\label{eq:staggeredAnsatz2}
\sigma^1_n \equiv m + s\cdot (-1)^{\sum\limits_{\mu} n_\mu}     
&\mbox{and}&
\lambda_n \equiv \lambda.
\eea
Minimizing $\hat S[\sigma^1,\lambda]$ with respect to $m$, $s$, and $\lambda$ will then yield the three gap equations
\bea
0&=&m\cdot \left[\lambda - \left(8\tilde\kappa_N \varphi_0^2 + \frac{16\rho^2}{\tilde y_N^2 \varphi_0^2} \cdot q(0)\right)  \right] \\
0&=&s\cdot \left[\lambda - \left(-8\tilde\kappa_N \varphi_0^2 + \frac{16\rho^2}{\tilde y_N^2 \varphi_0^2} \cdot  q\left(
\pi,\pi,\pi,\pi\right)  \right)  \right] \\
m^2+s^2 &=&  1 - \tilde \frac{t_N}{4}(1-\frac{1}{N}) \int\limits_{k\in\ImpSpace} \frac{d^4k}{(2\pi)^4}
\left[-\tilde\kappa_N \varphi_0^2 \sum\limits_{\mu}\cos(k_\mu) - \frac{8\rho^2}{\tilde y_N^2 \varphi_0^2} q(k) +
\frac{\lambda}{2}\right]^{-1} 
\eea
where $q(k)$ denotes the eigenvalues of $\left|\Gamma(\Delta x)\right|^2$ corresponding to plane waves 
and is given by
\bea
q(k) &=& \int\limits_{p\in\ImpSpace} \frac{d^4p}{(2\pi)^4}
\frac{\nu^+(p)}{\nu^+(p)-2\rho}\cdot \frac{\nu^+(\wp)}{\nu^+(\wp)-2\rho}
\cdot \frac{\tilde p\cdot \tilde \wp}{\sqrt{\tilde p^2}\cdot \sqrt{\tilde{\wp}^2}}, \quad \wp = k-p.
\eea
For the ferromagnetic phase, \ie $m\neq 0$ and $s=0$, one obtains a self-consistent determination equation for 
$m^2>0$ according to
\bea
0 < m^2 &=&  1 - \tilde \frac{t_N}{4}(1-\frac{1}{N}) \int\limits_{k\in\ImpSpace} \frac{d^4k}{(2\pi)^4}
\left[\tilde\kappa_N \varphi_0^2 \sum\limits_{\mu}\left(1-\cos(k_\mu)\right) 
+ \frac{8\rho^2}{\tilde y_N^2 \varphi_0^2} \left(q(0)-q(k)\right)\right]^{-1}.
\eea
An analogous relation holds for the antiferromagnetic phase (AFM). These relations can be treated numerically. 
The resulting phase diagram is shown in Figure~\ref{fig:PhaseDiagram2}. Again, we present a qualitative diagram only, 
since our results are preliminary. The appearance of three different phases, namely a symmetric, a ferromagnetic and
an antiferromagnetic phase, can be observed. The presented phase diagram is in qualitative agreement with earlier
results, using non-chiral fermions, see e.g.~\cite{Hasenfratz:1991it}.

\section{Summary and outlook}
In this paper we have studied the phase structure of a chiral invariant lattice Higgs-Yukawa model,
originally proposed by L\"uscher, by means of analytic $1/N_f$-expansions at tree-level. This was possible 
for small and for large values of the Yukawa coupling. Symmetric, ferromagnetic, and antiferromagnetic
phases have been observed in both regimes of the Yukawa coupling constant. Additionally, a ferrimagnetic
phase was found in the first regime. A next step would be the comparison of our analytical results
with corresponding Monte Carlo data.

\section*{Acknowledgements}
We thank the DFG for supporting this study by the DFG-project {\it Mu932/4-1}.
Furthermore, we would like to thank Michael M\"uller-Preussker for helpful
discussions and comments.

\nocite{*}
\bibliographystyle{unsrtOwnNoTitles}  
\bibliography{Proceedings}

\begin{thebibliography}{10}

\bibitem{Smit:1989tz}
J.~Smit.
\newblock Nucl. Phys. Proc. Suppl. 17, 3--16 (1990).

\bibitem{Shigemitsu:1991tc}
J.~Shigemitsu.
\newblock Nucl. Phys. Proc. Suppl. 20, 515--527 (1991).

\bibitem{Golterman:1990nx}
M.~F.~L. Golterman.
\newblock Nucl. Phys. Proc. Suppl. 20, 528--541 (1991).

\bibitem{book:Montvay}
I.~Montvay and G.~M{\"u}nster.
\newblock {C}ambridge U{}niversity {P}ress (1997).

\bibitem{book:Jersak}
A.~K. De and J.~Jers{\'a}k.
\newblock {HLRZ} {J\"u}lich, {HLRZ} 91-83, preprint edition (1991).

\bibitem{Golterman:1992ye}
M.~F.~L. Golterman, D.~N. Petcher, and E.~Rivas.
\newblock Nucl. Phys. Proc. Suppl. 29BC, 193--199 (1992).

\bibitem{Luscher:1998pq}
M.~L{\"u}scher.
\newblock Phys. Lett. B428, 342--345 (1998).

\bibitem{Neuberger:1998wv}
H.~Neuberger.
\newblock Phys. Lett. B427, 353--355 (1998).

\bibitem{Hasenfratz:1991it}
A.~Hasenfratz, P.~Hasenfratz, K.~Jansen, J.~Kuti, and Y.~Shen.
\newblock Nucl. Phys. B365, 79--97 (1991).

\bibitem{Hasenfratz:1992xs}
A.~Hasenfratz, K.~Jansen, and Y.~Shen.
\newblock Nucl. Phys. B394, 527--540 (1993).

\bibitem{Bock:1990tv}
Wolfgang Bock et~al.
\newblock Nucl. Phys. B344, 207--237 (1990).

\end{thebibliography}

\end{document}